\RequirePackage{fix-cm}
\documentclass[twocolumn,epjc3]{svjour3}
\smartqed  
\RequirePackage{graphicx}

\usepackage{amssymb,graphics}
\usepackage{epsfig,subfigure}
\usepackage{color}
\usepackage{amsmath}
\newcommand{\Lie}{\pounds}
\newcommand {\nn}    {\nonumber}
\journalname{Eur. Phys. J. C}
\begin{document}

\title{Time-Dependent Scalar Fields in Modified Gravities in a Stationary Spacetime}



\author{Yi Zhong\thanksref{e1}
        \and Bao-Ming Gu\thanksref{e2}
         \and Shao-Wen Wei\thanksref{e3}
         \and Yu-Xiao Liu\thanksref{e4}
        }

\thankstext{e1}{e-mail: zhongy13@lzu.edu.cn}
\thankstext{e2}{e-mail: gubm09@lzu.edu.cn}
\thankstext{e3}{e-mail: weishw@lzu.edu.cn}
\thankstext{e4}{e-mail: liuyx@lzu.edu.cn, corresponding author}


\institute{Institute of Theoretical Physics,
            Lanzhou University, Lanzhou 730000,
            People's Republic of China
}

\date{Received: date / Accepted: date}

\maketitle

\begin{abstract}
Most no-hair theorems involve the assumption that the scalar field
is independent of time.
Recently in [Phys. Rev. D90 (2014) 041501(R)] the existence of time-dependent scalar
 hair outside a stationary black hole in general relativity was ruled out.
 We generalize this work to modified gravities and non-minimally coupled scalar field with an additional assumption that the spacetime
 is axisymmetric. It is shown that
 in higher-order gravity such as metric $f(R)$ gravity the time-dependent scalar
 hair doesn't exist. While in Palatini $f(R)$ gravity and non-minimally
 coupled case the time-dependent scalar
 hair may exist.
\end{abstract}

\section{Introduction}

It is well known that black holes have no hair except the parameters of mass,
electric charge, and angular momentum. More precisely, the no-hair
theorem claims that all black hole solutions of the Einstein-Maxwell equations of gravitation and electromagnetism in general relativity can be completely characterized by only the three parameters. Since the long-range field in the standard model of
particle physics is electromagnetism, the
matter field considered in the original no-hair theorem is electromagnetic
field only. Nevertheless, it is still worth thinking about what the result is if we
take other matter fields { such as} scalar fields
into account.

The issue of the scalar-vacuum was first considered in 1970 in
Ref. \cite{raey1}.
 The canonical scalar hair was ruled out for scalar fields with
 various kinds of potential \cite{PhysRevLett.28.452,PhysRevD.5.1239,PhysRevD.5.2403}.
 Recently, this
proof was extended to non-canonical scalar fields
\cite{Graham:2014mda} and Galileons  \cite{Hui:2012qt,PhysRevD.90.124063}.
Besides, black holes in Brans-Dicke and
scalar-tensor theories of gravity were studied in Refs.
\cite{Hawking1972,PhysRevLett.108.081103}, which showed that the isolated
stationary black holes in scalar-tensor
theories of gravity are no different than in general relativity. In another
word, non-minimally coupled scalar hair is also ruled out for stationary and
conformally flat black holes. However, there is still the case that scalar
hair does
exists. Coexistence of black holes and a long-range scalar field in cosmology was
presented in Refs.~\cite{PhysRevLett.83.2699,Zloshchastiev:2004ny}. Other scalar hair cases can be found in Refs.
\cite{Martinez:2006an,Sotiriou:2013qea,PhysRevD.90.124063,Feng:2013tza,Liu:2013gja}.
These results are based on a same assumption: the
scalar field is time-independent.
In Ref. \cite{Zhang:2014sta}, the authors considered Einstein gravity minimally coupled to a dilaton scalar field and obtained an exact time-dependent spherically symmetric solution, which describes gravitational collapse to a static scalar-hairy black hole.
{ If the scalar field is time-dependent,
we should be more careful when defining the scalar hair in order to distinguish with some trivial situations. If a time-dependent scalar field is compatible with a stationary black hole metric (the back action of the scalar field to the spacetime is taken into account), it is called time-dependent scalar hair. Hence an in-falling flux of scalar waves is not time-independent scalar hair outside of a black hole, because the metric is no longer stationary if the back action to the spacetime is taken into account. It is important that the metric should be stationary because the no-hair theorem is about stationary black holes and the end state of the collapse of a star is stationary. It was shown that the system of a charged scalar field coupled to an electromagnetic field settles down to a stationary black hole with oscillating scalar hair \cite{Nicolas2015}.}
It has been shown that scalar fields do not necessarily share the symmetries with the spacetime \cite{Smolic:2015txa}.
In Ref. \cite{Graham:2014ina} it was shown
that the stationary spacetime does not ensure that the scalar field is
time-independent  and time-dependent real non-canonical scalar hair
was ruled out in Einstein gravity. While for the complex scalar field these arguments do not apply. Indeed Kerr black holes were found to
be having time-dependent
massive complex scalar hair \cite{Herdeiro:2014goa,Herdeiro:2015gia}.

In this paper, we would like to generalize the work of Ref.~\cite{Graham:2014ina} to
some modified gravities. Since the proof only needs a small subset of the
Einstein equations \cite{Graham:2014ina}, this generalization is turned out to be possible for some cases.
Among numerous modified gravities, $f(R)$ gravity, which is motivated by
high-energy physics, cosmology and astrophysics, has received increased
attention. It is interesting to consider the generalization of
Ref. \cite{Graham:2014ina} to $f(R)$ gravity. For metric $f(R)$ gravity
 the scalar curvature $R$ in the action is constructed from the metric only.
 And for Palatini $f(R)$ gravity
 the scalar curvature $R=g^{\mu\nu}R_{\mu\nu}$ where the Ricci curvature
 $R_{\mu\nu}$ is constructed from the independent connection.
 Besides, since the
time-independent non-minimally coupled scalar hair is ruled
out \cite{Hawking1972,PhysRevLett.108.081103}, we also investigate the case that the scalar field is
time-dependent and we will find nontrivial results.

This paper is organized as follows. In Sec. \ref{fRgravity}
we first investigate the time-dependent scalar field in metric
$f(R)$ gravity, then generalized it to other higher-order gravity and
Eddington Inspired Born-Infeld (EiBI) gravity \cite{PhysRevLett.105.011101}. In Sec. \ref{PalatinifRgravity} time-dependent scalar field in Palatini $f(R)$ gravity is investigated. In Sec. \ref{non-minimallyCoupled} we investigate the time-independent non-minimally coupled scalar.
Finally the conclusion is given in Sec. \ref{Conclusion}.

\section{Time-dependent scalar field in $f(R)$ gravity}
\label{fRgravity}

The action of $f(R)$ gravity is
\begin{eqnarray}
    S = \frac{1}{2\kappa}\int d^{4}x \sqrt{-g} f(R)
    + S_M (g_{\mu\nu},\varphi),
     \label{f(R) action}
\end{eqnarray}
where $\varphi$ denotes the matter field.
The variation of the action (\ref{f(R) action}) with respect to the
metric $g_{\mu\nu}$
leads to the equation of motion (EoM) in $f(R)$ gravity:
\begin{eqnarray}
    f_R R_{\mu\nu} - \frac{1}{2}f(R)g_{\mu\nu}
    + [ g_{\mu\nu}\Box - \nabla_{\mu}\nabla_{\nu} ] f_R
    = \kappa T_{\mu\nu},
    \label{f(R) EoM}
\end{eqnarray}
where $f_R \equiv \frac{\partial f(R)}{\partial R}$,
$\Box=\nabla^{\mu}\nabla_{\mu}$, and the energy-momentum tensor $T_{\mu\nu}$ is
given by
\begin{eqnarray}
    T_{\mu\nu} = -\frac{2}{\sqrt{-g}}\frac{\delta S_M}{\delta g^{\mu\nu}}.
\end{eqnarray}
In general relativity, if the null energy condition holds, the rigidity
theorem ensures that stationary spacetime must be axisymmetric
\cite{Hawking:1971vc,Hollands:2006rj}. In $f(R)$ gravity, the null energy
condition does not lead to $R_{\mu\nu}l^\mu l^\nu \geq 0$ for
all timelike vector $l^{\mu}$. Therefore, the null energy condition of the
matter fields does not lead to the conclusion that stationary spacetime
must be axisymmetric. We have to assume that the spacetime
is axisymmetric and we choose coordinates
$(t,r,\theta,\phi)$ so that the metric takes the
form \cite{chandrasekhar}
\begin{eqnarray}
    ds^2 =&& \!\!-e^{u(r,\theta)}dt^2 + 2\rho(r,\theta)dtd\phi+e^{v(r,\theta)}d\phi^2
    \nonumber\\
    &&\!\!\!\!\!\!\!\!\!+e^{A(r,\theta)}dr^2+e^{B(r,\theta)}d\theta^2.
     \label{metric}
\end{eqnarray}
One can easily verify that  the following components of the Ricci tensor
and Christoffel symbol vanish,
\begin{eqnarray}
   R_{tr}&=&R_{t\theta}=R_{r\phi}=R_{\theta\phi}=0,
   \label{vanish ricci}\\
   \Gamma^{r}_{tr}&=&\Gamma^{\theta}_{tr}
   =\Gamma^{r}_{t\theta}=\Gamma^{\theta}_{t\theta}=0.
   \label{vanish connection}
\end{eqnarray}

The action of the K-essence is
\cite{ArmendarizPicon:1999rj,PhysRevD.62.023511,PhysRevLett.85.4438,PhysRevD.63.103510}
\begin{eqnarray}
    S_{\text{M}} = \int d^{4}x \sqrt{-g} P(\varphi, X),
    \label{K action}
\end{eqnarray}
where the kinetic term is $X=-\frac{1}{2}\nabla_{\mu}\varphi\nabla^{\mu}\varphi$. When $P=X-V(\varphi)$,
Eq. (\ref{K action}) reduces to the action of a canonical scalar field.
Varying the
action (\ref{K action}) with respect to the scalar field $\varphi$ we obtain the
EoM of the non-canonical scalar field,
\begin{eqnarray}
    P_{\varphi}+P_X\Box\varphi +
    (\nabla^{\mu}\varphi)\nabla_{\mu}P_X=0.
    \label{K eom}
\end{eqnarray}
The
energy-momentum of the scalar field can be obtained by varying the
action (\ref{K action}) with respect to the metric:
\begin{eqnarray}
    T_{\mu\nu} = P_X
                 \partial_{\mu}\varphi\partial_{\nu}\varphi
                 + P g_{\mu\nu}.
\end{eqnarray}
the $tr$ and $t\theta$
components of Eq. (\ref{f(R) EoM}) imply that
\begin{eqnarray}
    T_{tr} &=& P_X
                 \partial_{t}\varphi\partial_{r}\varphi=0,
    \label{t,r}\\
    T_{t\theta} &=& P_X
                 \partial_{t}\varphi\partial_{\theta}\varphi=0.
    \label{t,theta}
\end{eqnarray}
Note that $P_X\neq0$, as otherwise the action
(\ref{K action}) would depend on $\varphi$ only and the EoM of
the scalar field would be an algebraic equation. Moreover, the scalar field is
time-dependent, i.e. $\partial_{t}\varphi\neq0$. Thus, Eqs.
(\ref{t,r}) and (\ref{t,theta}) yield
\begin{eqnarray}
    \partial_{r}\varphi = 0, ~~~~~~ \partial_{\theta}\varphi = 0,
\end{eqnarray}
or, equivalently,
\begin{eqnarray}
    \varphi = \varphi(t,\phi).
\end{eqnarray}
On the other hand, considering the $rr$ component of Eq. (\ref{f(R) EoM}) and
noting that the metric is independent of time, we have
\begin{eqnarray}
    \partial_t T_{rr} &=& \partial_t (P_X
                        \partial_{r}\varphi\partial_{r}\varphi)
                        + g_{rr}\partial_t P   \nonumber\\
                       &=&  g_{rr}\partial_t P = 0.
    \label{ptr}
\end{eqnarray}
Thus, $\partial_t P = 0$. Similarly, the $tt$ component of Eq. (\ref{f(R) EoM}) gives that $P_X \dot{\varphi}^2$ is independent of time. For general
actions these yield $P_\varphi=0$ and $\varphi$ depends at most linearly upon $\phi$, as otherwise there will be two equations for one unknown $\varphi$, the system will be overdetermined\cite{Graham:2014ina}.

The result that $P_\varphi=0$ was educed from a highbrow point of view in Ref. \cite{Smolic:2015txa}. Here we briefly introduce the derivation.
We start from a K-essence minimally coupled to gravity (not necessarily $f(R)$
gravity) in a stationary spacetime. It is easy to verify that
\begin{eqnarray}
T=g^{\mu\nu}T_{\mu\nu}=-2XP_X+4P,
\label{kt}
\end{eqnarray}
and
\begin{eqnarray}
P=\frac{1}{4}T \pm
\frac{1}{4}\sqrt{\frac{3T_{\mu\nu}T^{\mu\nu}-T^2}{3}}.
\label{kp}
\end{eqnarray}
Since the spacetime is stationary, from Eqs. (\ref{kt}) and (\ref{kp}) we have
\begin{eqnarray}
0 = \Lie_\xi P = P_{,X} \Lie_\xi X + P_{,\phi} \Lie_\xi \varphi,
\label{kp2}
\end{eqnarray}
and
\begin{eqnarray}
0 = \Lie_\xi T = -2 (\Lie_\xi X) P_{,X} - 2X \Lie_\xi (P_{,X}) + 4 \Lie_\xi P,
\label{kt2}
\end{eqnarray}
where $\xi$ is the time-like killing vector. Thus we have $\Lie_\xi (P_{,X}) = (\Lie_\xi P)_{,X} = 0$, which together with Eqs. (\ref{kp2}) and (\ref{kt2})
yields $P_{,\varphi} \Lie_\xi \varphi = 0$. So, the condition for the scalar field
not to inherit the symmetry of the spacetime is $P_{,\varphi}=0$.

Similarly one can deduce that the
scalar field $\varphi$ depends at most linearly upon $\phi$. Moreover,
since $\varphi$ should depend periodically upon $\phi$, it is incompatible
if $\varphi$ depends linearly upon $\phi$. Hence,
we finally deduce that the only possible configuration of
the scalar field is
\begin{eqnarray}
    \varphi = a t + b,
    \label{scalar func}
\end{eqnarray}
where $a$ and $b$ are constants. So far we have proved that in $f(R)$ gravity,
the time-dependent non-canonical scalar field in a stationary spacetime is
only a linear function of $t$. This conclusion is the same as that
of Ref. \cite{Graham:2014ina}. Following the procedure
of Ref. \cite{Graham:2014ina}, for asymptotically flat
and (anti-)de Sitter stationary black holes,
there is no time-dependent scalar hair. Here we give a brief demonstration.

\subsection{Boundary conditions}
Let's first consider the asymptotic flat condition,
i.e. $g_{\mu\nu} \rightarrow \eta_{\mu\nu}$ as the radial
coordinate $r \rightarrow \infty$,
for which $X \rightarrow a^2 /2$, and the $tt$ and $rr$
components of the energy-momentum tensor tend to
\begin{eqnarray}
    T_{tt}  &\rightarrow&  a^2 P_X (\frac{a^2}{2})
    - P(\frac{a^2}{2}),    \label{bdr11}\\
    T_{rr}  &\rightarrow&  P(\frac{a^2}{2}). \label{bdr12}
\end{eqnarray}
The EoMs demand $T_{tt}=0$ and $T_{rr}=0$, thus
either $a=0$ or $P_X (\frac{a^2}{2})$. However, $P_X (\frac{a^2}{2})$
together with $P(\frac{a^2}{2})$ compose two equations for one unknown,
which is overdetermined. Hence we have $a=0$, and the scalar field is a
constant.

For the case of an asymptotically anti-de Sitter spacetime,
$g^{tt}\rightarrow0$ as $r \rightarrow \infty$, which yields $X\rightarrow0$. In the static spherically symmetric
coordinates, the anti-de Sitter metric reads
\begin{eqnarray}
    ds^2  &=& -\Big(1-\frac{\Lambda}{3}r^2\Big)dt^2 + \Big(1-\frac{\Lambda}{3}r^2\Big)^{-1} dr^2
            \nonumber\\
            &+ & r^2 (d\theta^2 + \text{sin}^2 \theta d\phi^2).
    \label{ads metric}
\end{eqnarray}
The $tt$ and $rr$ components of the energy-momentum tensor at infinity tend to
\begin{eqnarray}
    T_{tt}  &\rightarrow&  P_X P(0)
    - \Big(1+\frac{|\Lambda|r^2}{3}\Big)P(0),    \\
    T_{tt}  &\rightarrow&  a^2 P_X(0)
    - \Big(1+\frac{|\Lambda|r^2}{3}\Big)P(0),    \\
    T_{rr}   &\rightarrow&  0.
\end{eqnarray}
It is clear that $T_{tt}=0$ and $T_{rr}=0$ yield $P(0)=0$ and $a=0$.
So there is no time-dependent scalar hair.

For the case of an asymptotically de Sitter spacetime, in the static
coordinates the metric is the same as Eq. (\ref{ads metric}). As $\Lambda>0$,
there is an event horizon at $r=\sqrt{3/\Lambda}$. Thus, $g_{rr} (r\rightarrow\sqrt{3/\Lambda})\rightarrow \infty$
 leads to $T_{rr}(r\rightarrow\sqrt{3/\Lambda})\rightarrow\infty$, which is incompatible
 with the geometry. Hence there is no time-dependent scalar hair.

The derivation of the time-dependent scalar field in $f(R)$ gravity can be
generalized to a large class of alternative  theories of gravity under the metric form (\ref{metric}).
As an example, consider a higher-order gravity with the action
\begin{eqnarray}
    S = \frac{1}{2\kappa}\int d^{4}x\sqrt{-g}( R  + \alpha R^{2}
        + \beta R_{\mu\nu}R^{\mu\nu}) + S_{\varphi}.
    \label{high action}
\end{eqnarray}
The field equations read
\begin{eqnarray}
    &&G_{\mu\nu}+2 \alpha R \Big(R_{{\mu\nu}}
           -\frac{1}{4}R ~g_{{\mu\nu}}\Big)
           \nonumber \\
    &+& (2\alpha + \beta ) ( g_{{\mu\nu}} \square
           - \nabla_{\mu}\nabla_{\nu} ) R \nonumber \\
    &+& 2 \beta R^{{\rho\sigma}} \Big(R_{{\mu\rho\nu\sigma}}
           - \frac{1}{4} R_{{\rho\sigma}}~g_{{\mu\nu}}\Big)\nonumber \\
   &+& \beta\square \Big(R_{{\mu\nu} }-\frac{1}{2} R~ g_{\mu\nu}\Big)
           = \kappa T^{\varphi}_{\mu\nu}.
\label{high eom}
\end{eqnarray}
Since the metric is stationary and axisymmetric, the $tr$ and $t\theta$
components of Eq. (\ref{high eom}) vanish and we still have
Eqs. (\ref{t,r}) and (\ref{t,theta}). The rest derivation is the same as that in
$f(R)$ gravity. It is obvious that these arguments can apply to some other alternative
gravities like  EiBI gravity theory.

\subsection{Double scalar fields}

We consider the case that the matter fields are consisted of two coupled non-canonical
scalar fields $\varphi_1$ and $\varphi_2$, of which the generalized action is
\begin{eqnarray}
    S_{\text{M}} = \int d^{4}x \sqrt{-g} P(\varphi_1, \varphi_2, X_1, X_2)
    \label{double Scalar action}.
\end{eqnarray}
This action contains the case of a complex scalar field as a special case.
The energy-momentum tensor is
\begin{eqnarray}
    T_{\mu\nu} = P_{X_1}
                 \partial_{\mu}\varphi_1 \partial_{\nu}\varphi_1
                 + P_{X_2}
                 \partial_{\mu}\varphi_1 \partial_{\nu}\varphi_2
                 + P g_{\mu\nu}.
   \label{double Scalar T}
\end{eqnarray}
Therefore $T_{0i}=0$ does not necessarily lead to
$\partial_{0}\varphi_1 \partial_{i}\varphi_1=0$
or $\partial_{0}\varphi_2 \partial_{i}\varphi_2=0$, and
the argument given above does not work for the double scalar field case any more.

\section{Time-dependent scalar field in Palatini $f(R)$ gravity}
\label{PalatinifRgravity}

Now we turn to the time-dependent scalar field in Palatini $f(R)$ gravity.
The action of Palatini $f(R)$ gravity is
\cite{Sotiriou:2006sr,Sotiriou:2006hs,Sotiriou:2008rp}
\begin{eqnarray}
\label{palaction}
 S_{\text{Pal}}=\frac{1}{2\kappa }\int d^4 x \sqrt{-g} \, f({\cal R})
+S_{\text{M}}(g_{\mu\nu}, \varphi),
\end{eqnarray}
where the Ricci tensor ${\cal R}_{\mu\nu}$ is
constructed with the independent connection $\Gamma^{\lambda}_{\mu\nu}$
and the corresponding Ricci scalar is ${\cal R}=g^{\mu\nu} {\cal R}_{\mu\nu}$.
Here we still assume that the spacetime is stationary and axisymmetric. Then
the metric has the same form of (\ref{metric}) and ${\cal R}$ is independent on
$t$ and $\phi$.

Varying the action~(\ref{palaction}) independently with respect
to the metric and connection, one can obtain the EoMs of Palatini
$f(R)$ gravity,
\begin{eqnarray}
f_{\cal R} {\cal R}_{(\mu\nu)}-\frac{1}{2}f({\cal
R})g_{\mu\nu}&=&\kappa \, T_{\mu\nu},
\label{ptneom11}\\
\widetilde{\nabla}_\lambda\left(\sqrt{-g}f_{\cal R}g^{\mu\nu}\right)&=&0,
\label{ptneom12}
\end{eqnarray}
where $\widetilde{\nabla}_\lambda$ is defined with the independent connection
$\Gamma^{\lambda}_{\mu\nu}$. Let us define a conformal metric $q_{\mu\nu}$,
\begin{eqnarray}
 \label{hgconf}
q_{\mu\nu} \equiv f_{\cal R}g_{\mu\nu}.
\end{eqnarray}
%
Then, Eq.~(\ref{ptneom12}) implies that the independent connection
$\Gamma^{\lambda}_{\mu\nu}$ is the Levi-Civita connection of the conformal
metric $q_{\mu\nu}$. 
Under conformal transformations, the Ricci
tensor $R_{\mu\nu}$ transforms as
\begin{eqnarray}
\label{confrel1}
{\cal R}_{\mu\nu} &=& R_{\mu\nu}+\frac{3}{2}
   \frac{1}{ f_{\cal{R}}^2   }\left(\nabla_\mu f_{\cal R}\right)\left(\nabla_\nu
f_{\cal R}\right) \nn\\
& -& \frac{1}{f_{\cal
R}}\left(\nabla_\mu
\nabla_\nu+\frac{1}{2}g_{\mu\nu}\Box\right)f_{\cal R},
\end{eqnarray}
where the Ricci tensor ${R}_{\mu\nu}$ and $\nabla_\mu$ are constructed by the spacetime metric $g_{\mu\nu}$.
Contraction with $g^{\mu\nu}$ yields
\begin{eqnarray}
\label{confrel2}
{\cal R}&=&R+\frac{3}{2f_{\cal R}^2}
    \left(\nabla_\mu f_{\cal{R}}  \right)
    \left(\nabla^\mu f_{\cal{R}}  \right)    \nn\\
 &-&\frac{3}{f_{\cal R}}\Box f_{\cal R}.
\end{eqnarray}
With Eqs. (\ref{confrel1}) and (\ref{confrel2}), Eq. (\ref{ptneom11})
is reduced to
\begin{eqnarray}
G_{\mu \nu} &= &\frac{\kappa}{f_{\cal{R}} }T_{\mu \nu}- \frac{1}{2}
g_{\mu \nu} \left({\cal R} - \frac{f}{ f_{\cal{R}} } \right)\nonumber \\
&+ &\frac{1}{ f_{\cal{R}} } \left(
			\nabla_{\mu} \nabla_{\nu}
			- g_{\mu \nu} \Box
		\right) f_{\cal{R}}  \nn\\
&- &\frac{3}{2}\frac{1}{ f_{\cal{R}}^2 } \left[
			(\nabla_{\mu}  f_{\cal{R}} )
            (\nabla_{\nu} f_{\cal{R}} )
			- \frac{1}{2}g_{\mu \nu} (\nabla f_{\cal{R}} )^2
		\right],
\label{ptneom2}
\end{eqnarray}
 from which we can see that we still have Eqs. (\ref{t,r})-(\ref{t,theta}) and Eqs. (\ref{ptr})-(\ref{kt2}) for Palatini $f(R)$ gravity.
Thus, the only possible configuration of the scalar field is (\ref{scalar func}).

Now we consider whether the configuration of the scalar field can be compatible
to the boundary conditions. First we consider the asymptotic flat boundary
condition. Note that the asymptotic flat boundary
condition implies that the metric $g_{\mu\nu}$ approaches the Minkowski
metric $\eta_{\mu\nu}$, while $q_{\mu\nu}$ approaches conformally Minkowski.
 Thus
Eqs. (\ref{bdr11}) and (\ref{bdr12}) no longer hold.
On the other hand, $\partial_t P=0$ and $\partial_t \varphi=0$ yield
$\partial_\varphi P=0$, and
$P_X=P_X({ a^2}/{2})$
is a constant. Thus $\nabla_{\mu}P_X=0$.
It is easy to verify $\Box \varphi=0$. Therefore, the  configuration of the
 scalar field of Eq. (\ref{scalar func}) is compatible with Eq. (\ref{K eom}),
the EoM of the scalar field.
The asymptotic
flat boundary condition no longer yields $a=0$. Similar argument
can be made in the asymptotic AdS/dS cases.   Hence for all the three kinds
of boundary conditions the time-dependent scalar hair may exist
 in Palatini $f(\mathcal{R})$ gravity.

Though the discussion above failed to exclude the time-dependent scalar hair for
an arbitrary Palatini $f({\cal R})$ gravity, it works if
\begin{eqnarray}
 f(\mathcal{R})= \mathcal{R} +\sum_{n=2}^{N} a_n \mathcal{R}^n.
 \label{ptnfraction2}
\end{eqnarray}
Taking the trace of Eq. (\ref{ptneom11}), we have
\begin{eqnarray}
f_{\cal R} {\cal R}-2f({\cal
R})&=&\kappa \, T.
\label{ptneom21}
\end{eqnarray}
Using Eq. (\ref{ptnfraction2}) and noting that $T_{\mu\nu}$ approaches a constant
at infinity, Eq. (\ref{ptneom21}) yields $\cal R = \text{constant}$. Considering the
asymptotic flat boundary
condition, Eq. (\ref{confrel2}) yields $\mathcal{R} = 0$, $f(0)=0$ and $f_\mathcal{R}(0)=1$.
Finally Eq. (\ref{ptneom2}) results in $T_{\mu\nu}=0$ at infinity and the time-dependent
scalar hair is ruled out. For the asymptotic AdS/dS we have the same conclusion.

 We can also investigate the  time-dependent scalar field $\varphi$ in scalar-tensor
gravity with the action
\begin{eqnarray}
\label{metactionH2}
S_{\text{st}}&=&\frac{1}{ 2\kappa }\int d^4 x \sqrt{-g} \Big[ U(\psi)
R-\frac{1}{2}h(\psi) \nabla_{\mu}\psi\nabla^{\mu}\psi
-V(\psi)\Big] \nonumber\\
 &+&S_{\text{M}}(g_{\mu\nu},\varphi).
\end{eqnarray}
where the action of the scalar field $\varphi$ is still given by Eq. (\ref{K action}).
Here we also assume that $g_{\mu\nu}$ is in the form of Eq. (\ref{metric}) and $\psi=\psi(r,\theta)$.
This action is equivalent to the action (\ref{f(R) action}) of metric $f(R)$ gravity
if $U(\psi)=\psi$ and $h(\psi)=0$
\cite{Chiba:2003ir,Flanagan:2004bz,Sotiriou:2006hs,Sotiriou:2008rp}, and
equivalent to the action (\ref{palaction}) of Palatini $f(R)$ gravity
if $U(\psi)=\psi$ and $h(\psi)=-\frac{3}{2\psi}$
\cite{Flanagan:2004bz,PhysRevD.72.083505,Sotiriou:2006hs,Sotiriou:2008rp}.
This case will be the same with that of time-dependent scalar field in
Palatini $f(\mathcal{R})$ gravity: we can educe the conclusion that the scalar field $\varphi$
 only depend linearly on $t$, but the boundary conditions do not exclude the
scalar hair, thus the scalar hair may exist. While for some specific $V(\psi)$ corresponding to the Lagrangian of palatini $f(\mathcal{R})$ gravity given by
Eq. (\ref{ptnfraction2}) we can
rule out the scalar hair.

\section{Non-minimally coupled scalar field}
\label{non-minimallyCoupled}

We now give the argument for a time-dependent non-minimally coupled
scalar field in a stationary spacetime. It should be note that
this is different from the case of the  time-dependent scalar field in scalar-tensor.
The action for the non-minimally coupled scalar field is
\begin{eqnarray}
S&=&\int d^4x \sqrt{-g} \Big[\varphi
R-\frac{\omega(\varphi)}{\varphi} \nabla^\mu \varphi
\nabla_\mu \varphi-V(\varphi)\Big]\,.
\label{staction}
\end{eqnarray}
By varying with respect to $g_{\mu\nu}$ and $\varphi$, one obtains the
field equations
\begin{eqnarray}
R_{\mu\nu}-\frac{1}{2} R g_{\mu\nu} & = &
\frac{\omega(\varphi)}{\varphi^2}\left(\nabla_\mu\varphi\nabla_\nu\varphi
-\frac{1}{2} g_{\mu\nu} \, \nabla^\lambda\varphi\nabla_\lambda\varphi\right)
\nonumber\\
&&\nonumber\\
 & + & \frac{1}{\varphi}\left(\nabla_\mu\nabla_\nu\varphi
-g_{\mu\nu}\Box\varphi\right)-\frac{V(\varphi)}{2\varphi}
g_{\mu\nu},
\label{steom1}\\
(2\omega+3)\Box\varphi  & = & -\omega_{\varphi}
\, \nabla^\lambda\varphi\nabla_\lambda\varphi+\varphi
 V_{\varphi}-2V  .
\label{steom2}
\end{eqnarray}

Here we still assume that the spacetime is stationary and axisymmetric.
Thus we have the metric (\ref{metric}) and
Eqs. (\ref{vanish ricci})-(\ref{vanish connection}). The $tr$ and $t\theta$
components of Eq. (\ref{f(R) EoM}) imply that
\begin{eqnarray}
   \frac{\omega(\varphi)}{\varphi^2}
                 \partial_{t}\varphi\partial_{r}\varphi
                 +\frac{1}{\varphi}\nabla_t \nabla_r\varphi&=&0,
    \label{stt,r}\\
     \frac{\omega(\varphi)}{\varphi^2}
                 \partial_{t}\varphi\partial_{\theta}\varphi
                  +\frac{1}{\varphi}\nabla_t \nabla_\theta\varphi&=&0.
    \label{stt,theta}
\end{eqnarray}
Now it is clear that $\partial_{t}\varphi\neq0$ no longer educes
$\partial_{r}\varphi=0$ or $\partial_{\theta}\varphi=0$, thus the arguments
in Sec.
\ref{fRgravity} no longer apply and the time-dependent non-minimally coupled scalar hair
may exist.

\section{Conclusion}
\label{Conclusion}

In this paper, we investigated the non-canonical time-dependent scalar field
in a stationary and axisymmetric spacetime in modified gravities.
For a single real scalar field in metric $f(R)$ gravity,
we proved that the time-dependent
scalar hair does not exist for the
three kinds of boundary conditions (asymptotically flat, anti-de Sitter, and de Sitter).
It was shown that the demonstration can be
generalized to a large class of alternative  theories of gravity like the higher-order gravity described by the action (\ref{high action}) and EiBI gravity. While for two coupled scalar fields,
these arguments do not apply.
These conclusions are the same as the time-dependent scalar field in general relativity in Ref.\cite{Graham:2014ina}.

Though the demonstrations for a single scalar hair in general relativity and metric $f(R)$
only use a small subset of the field equations \cite{Graham:2014ina}, the generalization to other alternative gravities may not be correct.
For Palatini $f(R)$ gravity coupled with a scalar field, as the boundary conditions no longer ruled out the non-trivial configuration of the scalar field,
the time-dependent scalar hair may exist outside a stationary and axisymmetric black hole.

However, for the case that $f(\mathcal{R})$ is given by Eq. (\ref{ptnfraction2}),
the time-dependent scalar hair is ruled out. This conclusion can be
 widely generalized. The keypoint is to ensure that
$f(\mathcal{R}=0)=0$ and $f_{\mathcal{R}}(\mathcal{R}=0)\neq0$.

Since Palatini $f(R)$ gravity is equivalent to scalar-tensor gravity, similar
argument can be applied to time-dependent scalar field in scalar-tensor
gravity. For some specific $V(\psi)$ we can
rule out the scalar hair.
 For the time-independent
non-minimally coupled scalar field, since the effective energy-momentum of the
scalar field contains the second derivative of the scalar field, the derivations
of Sec. \ref{fRgravity} do not apply any more and non-minimally coupled scalar hair may exist outside a stationary and axisymmetric black hole.

\section*{Acknowledgement}

{ We would like to thank the referee for helping us to clarify the concept of the black hole scalar hair in the time-dependent case.}
This work was supported by the National Natural Science Foundation of
China (Grants No. 11205074, No. 11375075, { and No. 11522541}) and the Fundamental Research Funds for the Central Universities (Grants No. lzujbky-2015-jl1 and No. lzujbky-2015-207).


\begin{thebibliography}{10}
\providecommand{\url}[1]{{#1}}
\providecommand{\urlprefix}{URL }
\expandafter\ifx\csname urlstyle\endcsname\relax
  \providecommand{\doi}[1]{DOI \discretionary{}{}{}#1}\else
  \providecommand{\doi}{DOI \discretionary{}{}{}\begingroup
  \urlstyle{rm}\Url}\fi

\bibitem{raey1}
J.~Chase, Commun. Math. Phys. \textbf{19}(4), 276 (1970).
\newblock \doi{10.1007/BF01646635}.

\bibitem{PhysRevLett.28.452}
J.D. Bekenstein, Phys. Rev. Lett. \textbf{28}, 452 (1972).
\newblock \doi{10.1103/PhysRevLett.28.452}.

\bibitem{PhysRevD.5.1239}
J.D. Bekenstein, Phys. Rev. D \textbf{5}, 1239 (1972).
\newblock \doi{10.1103/PhysRevD.5.1239}.

\bibitem{PhysRevD.5.2403}
J.D. Bekenstein, Phys. Rev. D \textbf{5}, 2403 (1972).
\newblock \doi{10.1103/PhysRevD.5.2403}.

\bibitem{Graham:2014mda}
A.A.H. Graham, R.~Jha, Phys. Rev. D \textbf{89}, 084056 (2014).
\newblock \doi{10.1103/PhysRevD.89.084056}

\bibitem{Hui:2012qt}
L.~Hui, A.~Nicolis, Phys. Rev. Lett. \textbf{110}(24), 241104 (2013).
\newblock \doi{10.1103/PhysRevLett.110.241104}

\bibitem{PhysRevD.90.124063}
T.P. Sotiriou, S.Y. Zhou, Phys. Rev. D \textbf{90}, 124063 (2014).
\newblock \doi{10.1103/PhysRevD.90.124063}.

\bibitem{Hawking1972}
S.~Hawking, Commun. Math. Phys. \textbf{25}(2), 167 (1972).
\newblock \doi{10.1007/BF01877518}.

\bibitem{PhysRevLett.108.081103}
T.P. Sotiriou, V.~Faraoni, Phys. Rev. Lett. \textbf{108}, 081103 (2012).
\newblock \doi{10.1103/PhysRevLett.108.081103}.

\bibitem{PhysRevLett.83.2699}
T.~Jacobson, Phys. Rev. Lett. \textbf{83}, 2699 (1999).
\newblock \doi{10.1103/PhysRevLett.83.2699}.

\bibitem{Zloshchastiev:2004ny}
K.G. Zloshchastiev, Phys. Rev. Lett. \textbf{94}, 121101 (2005).
\newblock \doi{10.1103/PhysRevLett.94.121101}

\bibitem{Martinez:2006an}
C.~Martinez, R.~Troncoso, Phys. Rev. D \textbf{74}, 064007 (2006).
\newblock \doi{10.1103/PhysRevD.74.064007}

\bibitem{Sotiriou:2013qea}
T.P. Sotiriou, S.Y. Zhou, Phys. Rev. Lett. \textbf{112}, 251102 (2014).
\newblock \doi{10.1103/PhysRevLett.112.251102}

\bibitem{Feng:2013tza}
X.H. Feng, H.~Lu, Q.~Wen, Phys. Rev. D \textbf{89}, 044014 (2014).
\newblock \doi{10.1103/PhysRevD.89.044014}

\bibitem{Liu:2013gja}
H.S. Liu, H.~L¨¹, Phys. Lett. B \textbf{730}, 267 (2014).
\newblock \doi{10.1016/j.physletb.2014.01.056}

\bibitem{Zhang:2014sta}
X.~Zhang, H.~Lu, Phys. Lett. B \textbf{736}, 455 (2014).
\newblock \doi{10.1016/j.physletb.2014.07.052}


\bibitem{Nicolas2015}
N. Sanchis-Gual, J. C. Degollado, P. J. Montero, J. A. Font, C. Herdeiro, arXiv:1512.05358 [gr-qc]  (2015)

\bibitem{Smolic:2015txa}
I.~Smoli\'c, arXiv:1501.04967 [gr-qc]  (2015)

\bibitem{Graham:2014ina}
A.A.H. Graham, R.~Jha, Phys. Rev. D \textbf{90}, 041501 (2014).
\newblock \doi{10.1103/PhysRevD.90.041501}

\bibitem{Herdeiro:2014goa}
C.A.R. Herdeiro, E.~Radu, Phys. Rev. Lett. \textbf{112}, 221101 (2014).
\newblock \doi{10.1103/PhysRevLett.112.221101}

\bibitem{Herdeiro:2015gia}
C.~Herdeiro, E.~Radu, arXiv:1501.04319 [gr-qc]  (2015)

\bibitem{PhysRevLett.105.011101}
M.~Ba\~nados, P.G. Ferreira, Phys. Rev. Lett. \textbf{105}, 011101 (2010).
\newblock \doi{10.1103/PhysRevLett.105.011101}.

\bibitem{Hawking:1971vc}
S.~Hawking, Commun. Math. Phys. \textbf{25}, 152 (1972).
\newblock \doi{10.1007/BF01877517}

\bibitem{Hollands:2006rj}
S.~Hollands, A.~Ishibashi, R.M. Wald, Commun. Math. Phys. \textbf{271}, 699
  (2007).
\newblock \doi{10.1007/s00220-007-0216-4}

\bibitem{chandrasekhar}
S.~Chandrasekhar, \emph{The Mathematical Theory of Black Holes} (Oxford: Oxford
  UP, 1983)

\bibitem{ArmendarizPicon:1999rj}
C.~Armendariz-Picon, T.~Damour, V.F. Mukhanov, Phys. Lett. B \textbf{458}, 209
  (1999).
\newblock \doi{10.1016/S0370-2693(99)00603-6}

\bibitem{PhysRevD.62.023511}
T.~Chiba, T.~Okabe, M.~Yamaguchi, Phys. Rev. D \textbf{62}, 023511 (2000).
\newblock \doi{10.1103/PhysRevD.62.023511}.

\bibitem{PhysRevLett.85.4438}
C.~Armendariz-Picon, V.~Mukhanov, P.J. Steinhardt, Phys. Rev. Lett.
  \textbf{85}, 4438 (2000).
\newblock \doi{10.1103/PhysRevLett.85.4438}.

\bibitem{PhysRevD.63.103510}
C.~Armendariz-Picon, V.~Mukhanov, P.J. Steinhardt, Phys. Rev. D \textbf{63},
  103510 (2001).
\newblock \doi{10.1103/PhysRevD.63.103510}.

\bibitem{Sotiriou:2006sr}
T.P. Sotiriou,  pp. 1223--1225 (2006).
\newblock \doi{10.1142/9789812834300-0112}

\bibitem{Sotiriou:2006hs}
T.P. Sotiriou, Class. Quant. Grav. \textbf{23}, 5117 (2006).
\newblock \doi{10.1088/0264-9381/23/17/003}

\bibitem{Sotiriou:2008rp}
T.P. Sotiriou, V.~Faraoni, Rev. Mod. Phys. \textbf{82}, 451 (2010).
\newblock \doi{10.1103/RevModPhys.82.451}

\bibitem{Chiba:2003ir}
T.~Chiba, Phys. Lett. B \textbf{575}, 1 (2003).
\newblock \doi{10.1016/j.physletb.2003.09.033}

\bibitem{Flanagan:2004bz}
E.E. Flanagan, Class. Quant. Grav. \textbf{21}, 3817 (2004).
\newblock \doi{10.1088/0264-9381/21/15/N02}

\bibitem{PhysRevD.72.083505}
G.J. Olmo, Phys. Rev. D \textbf{72}, 083505 (2005).
\newblock \doi{10.1103/PhysRevD.72.083505}.


\end{thebibliography}

\end{document}